\documentstyle[12pt]{article}
\begin{document}
\begin{center}
{\large {\bf Lagrangian approach to integrable systems\\ yields
new symplectic structures for KdV }}\\[20mm] {\large {\bf Y.
Nutku}}\\[2mm] Feza G\"ursey Institute, \c{C}engelk\"oy 81220
Istanbul, Turkey\\ November 3, 2000
\\[15mm]
\end{center}

In the literature on integrable systems we find Hamiltonian
operators without explanation. There is a notable silence on
Lagrangians from which these Hamiltonian and symplectic structures
can be derived. We show that starting with Lagrangians, which turn
out to be degenerate, the Hamiltonian operators for integrable
systems can be constructed using Dirac's theory of constraints. We
illustrate this by giving a systematic discussion of the first
Hamiltonian structure of KdV. First by Dirac's theory and then
applying the covariant Witten-Zuckerman theory of symplectic
structure we arrive at its flux. Then we turn to a new Lagrangian
for KdV recently obtained by Pavlov and derive the corresponding
new symplectic structure for KdV. We show that KdV admits
infinitely many Lagrangian formulations and therefore infinitely
many symplectic structures.

\section{Introduction}

While there is no precise definition of complete integrability
there are many properties that we expect from a completely
integrable system which in fact enable us to recognize it as such.
Bi-Hamiltonian structure for which we have the celebrated theorem
of Magri \cite{magri} is a case in point. Non-linear evolution
equations that can be cast into Hamiltonian form in two
inequivalent but compatible ways admit a full set of conserved
quantities required by complete integrability. These conserved
Hamiltonians are in involution with respect to Poisson brackets
defined by both Hamiltonian structures.

We recall from earliest school that Hamiltonian structure is
derivable from a Lagrangian, yet in complete integrability there
is a notable, even deafening silence on Lagrangians. Why?

It turns out that in the variational formulation of integrable
equations the Lagrangians are degenerate and therefore require the
use of Dirac's theory of constraints \cite{dirac} in order to
achieve Hamiltonian form. On the other hand, degenerate
Lagrangians for integrable systems always admit a complete set of
second class constraints which enables us to eliminate all the
momenta and write the Hamiltonian equations of motion solely in
terms of the original variables. In effect this means that we can
often guess the Dirac bracket without recourse to Dirac's theory.
Thus in the literature on integrable systems Dirac's theory of
constraints hardly even gets a nod while Dirac brackets are
generally known as Hamiltonian operators. Even in the excellent
book of Dorfman \cite{irene} that she was able to complete before
her untimely death, Dirac's name appears in the title but in its
pages no reference can be found to a single degenerate Lagrangian
for which the Hamiltonian structure is obtained through Dirac's
theory of constraints. So, for both reasons, namely the fact that
Dirac's theory is not in the arsenal of most mathematicians as
well as the convenience afforded by second class constraints of
eliminating the momenta completely, the present literature
contains very little on degenerate Lagrangians for integrable
systems. It was, however, my original motivation for going into
integrable systems \cite{yavuz}.

Inevitably something as fundamental as the Lagrangian would make a
spectacular come-back and indeed it has. The covariant approach to
symplectic structure which has relatively recently been formulated
by Witten \cite{witten} and Zuckerman \cite{zuck} employs the
Lagrangian as its head-piece.

The usual Hamiltonian approach is not covariant because it singles
out an independent variable as ``time" with respect to which the
evolutionary system will be defined. Dirac was well aware of this
short-coming because he was ultimately interested in the
Hamiltonian formulation of Einstein's general relativity
\cite{diracadm} and in his book \cite{dirac} complains several
times about the fact that the Hamiltonian formalism is not
covariant. The Witten-Zuckerman covariant theory of symplectic
structure is very much in the spirit of Dirac's work. It starts
with a Lagrangian and in the construction of the symplectic
$2$-form the most crucial role is played by the boundary terms in
the first variation of the action. We shall now illustrate the
prominence of the Lagrangian in the covariant theory of symplectic
structure by applying the Witten-Zuckerman construction of the
symplectic $2$-form to the grandmother of all integrable systems,
namely KdV. We shall show that not only do we recover some
familiar results but we shall also able to present some new
symplectic structures starting from new Lagrangians for the KdV
equation. Shockingly enough, we shall find that there are still
unknown fundamental  results in the theory of KdV.

\section{KdV as bi-Hamiltonian system}

The bi-Hamiltonian structure of KdV was the first one to be
discovered and has consequently served as the model for the
multi-Hamiltonian structure of all integrable systems. It is well
known that the KdV equation
\begin{equation}
u_{t} + 6 \, u  \, u_x + u_{xxx} = 0 \label{kdv}
\end{equation}
can be cast into the form of Hamilton's equations
\begin{equation}
u_{\,t} = {\bf X} ( u )  = \{ u , H \}_D = J \; \delta_u H
\label{hameq}
\end{equation}
where $ {\bf X}$ is the vector field defining the flow which is
manifest from eq.(\ref{kdv}), $J$ is the Hamiltonian operator
defining the Poisson bracket and $\delta_u$ denotes $\delta /
\delta u$, the variational derivative with respect to $u$. The
Hamiltonian operator which is a skew-symmetric matrix of
differential operators satisfying the Jacobi identities is simply
obtained from the Dirac bracket as the subscript $D$ in
eq.(\ref{hameq}) indicates. However, this was not the historical
route to the Hamiltonian formulation of KdV which started with an
important paper by Gardner \cite{gardner} and independently by
Zakharov and Fadeev \cite{zf} where they showed that the infinite
set of conserved Hamiltonians $H_i, i=1, \dots, \infty$ commute
with respect to Poisson brackets defined by
\begin{equation}
 \left\{ H_i , \frac{d}{d x} \, H_k \right\}_P = 0,
\label{comm}
\end{equation}
which, as we shall soon show explicitly, is of course a particular
example of the Dirac bracket. Thus
\begin{equation}
J_0 = \frac{d}{d x}    \label{gardner}
\end{equation}
is the first Hamiltonian operator for KdV. The next important
development was an unpublished but widely known result of Lenard,
namely the recursion operator for KdV. Magri \cite{magri} realized
that it leads to the second Hamiltonian operator for KdV
\begin{equation}
J_1 = \frac{d^3}{d x^3} + 2 u  \, \frac{d}{d x} + \frac{d}{d x}
\, 2 u \label{magri}
\end{equation}
and was able to formulate his theorem on bi-Hamiltonian structure.
Lenard's recursion operator is simply
\begin{equation}
{\cal R} = J_1 \; (J_0)^{-1} \label{recursionop}
\end{equation}
and in the bi-Hamiltonian formulation of KdV the recursion
relation for conserved Hamiltonians is given by
\begin{equation}
u_{\,t_n} = J_0 \; \delta_u H_{n+1}  = J_1 \; \delta_u H_{n}
\label{lenard}
\end{equation}
with Hamiltonian densities given by
\begin{eqnarray} {\cal H}_0 & = & \frac{1}{2} u^2, \label{h0} \\
{\cal H}_1 & = & u^3-\frac{1}{2}u_x^{\;2}\label{h1}\\
{\cal H}_2 & = & \frac{5}{2} \, u^4 - 5 \, u \, u_x^{\;2}
+ \, \frac{1}{2} \, u_{xx}^{\;\;\;2}\label{h2}\\ .. & & ... \nonumber
\end{eqnarray}
which determines the higher flows in the KdV hierarchy.

In effect, there was no derivation of these
Hamiltonian operators. When we are presented with a result like
this all we can do is to check the properties of skew-symmetry and
Jacobi identities required of a Poisson bracket. Then we must keep
quiet, but it is an uneasy quiet.

\section{Dirac bracket is the Hamiltonian operator}

The first correct identification of the Hamiltonian operator
(\ref{gardner}) as the Dirac bracket appeared in a paper by
Macfarlane \cite{mcf} which, as far as I know, is unpublished and
has not been given the attention that it deserves. The first
systematic application of Dirac's theory of constraints to the
degenerate Lagrangian for KdV was given in \cite{ynkdv}.

We start with the variational formulation of KdV. For this purpose
we need to introduce
\begin{equation}
u = \phi_x, \label{clebsch}
\end{equation}
the Clebsch velocity potential. The potential KdV is
\begin{equation}
\phi_{t} + 3 \, \phi_{x}^{\;2} + \phi_{xxx} = 0 \label{potkdv}
\end{equation}
and it can be directly verified that the equations of motion following
from the variational principle
\begin{equation}
\delta I = 0\, ,   \; \;\;\;\;\;\; I = \int {\cal L} \; d t \, d x
\label{varpr}
\end{equation}
with Lagrangian density
\begin{equation}
{\cal L}_0 = \frac{1}{2}\, \phi_t \, \phi_x + \phi_x^{\,3} -
\frac{1}{2} \,  \phi_{xx}^{\;\;2} \label{lag1}
\end{equation}
yield KdV. This Lagrangian is degenerate because its Hessian
\begin{equation}
\frac{\partial^2 {\cal L} }{\partial \phi_{t}^{\;2} } = 0
\label{hessian}
\end{equation}
vanishes identically. Alternatively, the canonical momentum given
by
\begin{equation}
\pi =  \frac{\partial {\cal L}_0 }{\partial \phi_t} =
\frac{1}{2}\, \phi_{x} \label{momenta}
\end{equation}
cannot be inverted for the velocity $\phi_t$ and we have a
degenerate Lagrangian system. Therefore the Hamiltonian
formulation of the Lagrangian (\ref{lag1}) requires the use of
Dirac's theory of constraints.

Following Dirac \cite{dirac} we introduce the primary constraint
that results from the definition of the momentum (\ref{momenta})
\begin{equation}
\Phi =  \pi - \frac{1}{2} \, \phi_x  \label{constraint1}
\end{equation}
and {\it a priori} require that it vanish either ``weakly," or as
a ``strong" equation depending on whether or not the constraint is
first, or second class in Dirac's terminology. This is determined
by evaluating the Poisson bracket of the constraints. If the
result should vanish modulo the constraints then it is a first,
otherwise second class constraint. A second class constraint is an
equation that can be solved to eliminate a canonical variable.

In order to calculate the Poisson bracket of the constraints we
use the canonical Poisson bracket relations
\begin{equation}
 \{ \pi (x) , \phi (y) \}_P  =  \delta ( x - y)
\label{canonical}
\end{equation}
with all others vanishing. The result
\begin{equation}
 \{\Phi(x), \Phi(y)\}_P   =  \frac{1}{2}  \delta_{y} (x-y)
      - \frac{1}{2} \delta_{x} (y-x)
\label{pbconst}
\end{equation}
shows that the constraint (\ref{constraint1}) is second class as
it does not vanish by virtue of the constraint itself. A word of
caution, that I owe to my friend Galv\~{a}o \cite{capg}, is in
order here: One should not simplify the Poisson brackets of the
constraints (\ref{pbconst}) using the rules for manipulating
distributions at intermediate stages of calculation. In our case
we shall need to invert this distribution and therefore we must
keep the cumbersome appearance of eq.(\ref{pbconst}).
Simplifications should be made only at final results.

     The total Hamiltonian of Dirac is given by
\begin{eqnarray}
   H_{T} & = & \displaystyle{\int}
\left( \pi \, \phi_t - \displaystyle{\cal L}  + \lambda \, \Phi
\right) d x  \label{ht}  \\
 & = & \int \left[ - \phi_{x}^{\;3} + \frac{1}{2} \, \phi_{xx}^{\;\;\;2}
+ \lambda \left(  \pi - \frac{1}{2} \phi_x \right) \, \right] d x
\label{ht1}
\end{eqnarray}
where $\lambda$ is a Lagrange multiplier. The condition that the
constraint is maintained in time
\begin{equation}
  \{ \Phi (x) , H_{T} \} = 0
\label{seco}
\end{equation}
gives rise to no further constraints which would have been
secondary constraints. Instead, using eqs.(\ref{pbconst}), we find
that the Lagrange multiplier is completely determined from
eq.(\ref{seco})
\begin{equation}
\lambda = - 3 \, \phi_{x}^{\;2} - \phi_{xxx} \label{multiplier}
\end{equation}
which is expected, since the constraint and therefore the total
Hamiltonian is linear in the momenta the correct equation of
motion (\ref{potkdv}) will result only if the Lagrange multiplier
is simply the component of the vector field defining the flow for
potential KdV. The total Hamiltonian density of Dirac
\begin{equation}
   {\cal H}_{T} =  \frac{1}{2} \, \phi_{x}^{\;3}
   - \pi  \left(3\,\phi_{x}^{\;2}+\phi_{xxx}  \right)
\label{ht2}
\end{equation}
follows from the substitution of the Lagrange multiplier
(\ref{multiplier}) in eq.(\ref{ht1}). Now the check that all the
Hamiltonian equations of motion are satisfied with the Hamiltonian
(\ref{ht2}) is straight-forward. We can summarize all of them in
Hamilton's equations
\begin{equation}
{\cal A}_{t} = \{ {\cal A} , H_{T} \} \label{eqsofmo}
\end{equation}
where ${\cal A}$ is any smooth functional of the canonical
variables $\phi, \pi$ and their derivatives.

There is, however, one further and very important simplification
that we can carry out because in Dirac's theory second class
constraints hold as strong equations. This fact enables us to
eliminate the momentum in the total Hamiltonian (\ref{ht2}) using
the solution of eq.(\ref{momenta}). Thus we find
\begin{equation}
{\cal H}_{T} = - \, \phi_{x}^{\;3} + \frac{1}{2} \, \phi_{xx}^{\;\;\;2}
\end{equation}
for Dirac's total Hamiltonian density. Apart from an overall minus
sign, this is just the Hamiltonian function (\ref{h1}) for the
first Hamiltonian structure of KdV.

The Dirac bracket is the projection of the Poisson bracket from
phase space onto the hyper-surface defined by the constraint.
Given any two differentiable functionals of the canonical variables
${\cal A}$ and ${\cal B}$, the Dirac bracket is defined by
\begin{equation} \begin{array}{ll}
\{ {\cal A}(x), {\cal B}(y) \}_{D} = & \{ {\cal A}(x), {\cal B}(y) \} \\[4mm]
      & - \, \displaystyle{\int} \{ {\cal A}(x), \Phi(z) \}
J(z,w)  \{ \Phi (w) , {\cal B}(y) \} \, d z \, d w
\end{array}
\label{defdirac}
\end{equation}
where $J$ is the inverse of the matrix of Poisson brackets of the
constraints. The definition of the inverse is simply
\begin{equation}
\int \{\Phi (x), \Phi (z) \} J (z,y) \; d z
                  =  \delta ( x - y )
\label{mik}
\end{equation}
which results in a differential equation to be solved for $J$.
Starting with the Poisson bracket relation (\ref{pbconst}) we find
that eq.(\ref{mik}) can be solved readily to yield
\begin{eqnarray}
 J (x, y) & \equiv & J(x) \delta (x-y) \label{defhop}\\
& = & \left(\frac{d}{d x}\right)^{-1} \delta (x-y) = \theta (x-y)
\label{mm}
\end{eqnarray}
where $\theta$ is the Heaviside unit step function and $\left( d /
d x \right)^{-1}$ is the principal value integral \cite{fokas}.
With the definition of the Hamiltonian operator $J(x)$ given above
this principal value integral is the first Hamiltonian operator of
KdV in terms of the Clebsch potential.

Now we need to make contact with the first Hamiltonian operator
(\ref{gardner}) for KdV which is expressed in terms of the
velocity field $u$ rather than its potential $\phi$. In view of
the definition of the potential (\ref{clebsch}) and the
transformation properties of Hamiltonian operators, we find
\begin{equation}
 J_{u(x)} =  \frac{d}{d x} \;\; \left(\frac{d}{d x}\right)^{-1}
 \, \frac{d}{d x}  = \frac{d}{d x}
\label{gardneru}
\end{equation}
which is the same as (\ref{gardner}). This is the derivation of
the first Hamiltonian operator for KdV from first principles, {\it
i.e.} from a Lagrangian approach. I have gone into it in some
detail not only because it is a nice illustrative example but also
because carrying out the same construction for the second
Hamiltonian operator (\ref{magri}) is a highly non-trivial
unsolved problem.

\section{Symplectic form of KdV}
\label{sec-symp}

The Hamiltonian operator maps differentials of functions into
vector fields and in the opposite direction we have the symplectic
$2$-form $\omega$ which is the principal geometrical object in the
theory of symplectic structure. The symplectic $2$-form is closed
\begin{equation}
\delta \omega  = 0                      \, , \label{closed}
\end{equation}
which is equivalent to the Jacobi identities satisfied by the
Dirac bracket. By Poincar\'e's lemma, in a local neighborhood,
$\omega$ can be written as
\begin{equation}
\omega =  \delta  \alpha, \label{alphao}
\end{equation}
where $\alpha$ is a $1$-form. The statement of the symplectic
structure of the equations of motion consists of
\begin{equation}
 i_{\bf X} \omega  =  \delta H   \label{symhameq}
\end{equation}
which is obtained by the contraction of the symplectic 2-form
$\omega$ with the vector field {\bf X} defining the flow.

The symplectic 2-form is given by
\begin{equation}
\omega = \frac{1}{2} \int \delta \phi(x) \wedge K(x,y) \, \delta
\phi(y) \; d y \label{defomega}
\end{equation}
where $K(x,y)$ is the inverse of $J(x,y)$. This route of arriving at the
symplectic $2$-form through the Hamiltonian operator is
commonplace in the literature of integrable systems. But in the
Lagrangian approach it is absurd because we have the answer already
in eq.(\ref{mik}): Hamiltonian operators are Dirac brackets for
systems subject to second class primary constraints which are
obtained by {\it inverting} the Poisson bracket of the
constraints! Hence
\begin{equation}
 K( x, y )  \equiv \{ \Phi (x) , \Phi (y) \} ,
\label{pbceqinv}
\end{equation}
the symplectic $2$-form density can be obtained directly
from the Poisson bracket of second class constraints. In the case
of the first Hamiltonian structure of KdV from eq.(\ref{pbconst})
we find
\begin{equation}
 \omega_0 = \delta \phi \wedge \delta \phi_x
\label{om1}
\end{equation}
for the first symplectic $2$-form of KdV.

It remains to check Hamilton's equations in the symplectic form
(\ref{symhameq}) for KdV using this symplectic $2$-form and $H_1$.
For this purpose we need to contract the $2$-form (\ref{om1}) with
the vector field
\begin{equation}
{\bf X} = - \left( 3 \, \phi_x^{\;2} + \phi_{xxx} \right) \,
\frac{\delta}{\delta \phi} \label{vectorfield}
\end{equation}
defining the flow for potential KdV. Recalling
\begin{equation}
  i_{\delta / \delta \phi(y)}  \delta \phi_x = \delta_x (x-y),
\label{contraction}
\end{equation}
it follows immediately that eq.(\ref{symhameq}) is satisfied with
$H_1$ given by (\ref{h1}).

\section{Witten-Zuckerman 2-form}
\label{sec-wz}

Time plays a privileged role in Hamiltonian mechanics. While this
presents no problem for systems with finitely many degrees of
freedom, in field theory it has the severe disadvantage of
non-covariance. In order to remedy this situation Witten
\cite{witten} and Zuckerman \cite{zuck} have introduced the
conserved current 2-form which provides an elegant covariant
formulation of symplectic structure. The Witten-Zuckerman 2-form
$\omega^{\mu}$, where in our case $\mu$ ranges over two values $t$
and $x$, satisfies
\begin{eqnarray}
\delta \omega^{\mu} & = & 0, \label{wz2fclosed}\\
\omega^{\mu}_{\;,\mu} & = & 0, \label{wz2fcons}
\end{eqnarray}
the properties of closure and conservation. Hamilton's equations
can now be written in the covariant form
\begin{equation}
 i_{\bf X} \, \omega^\mu  =  \delta H^\mu   \label{symhameqcov}
\end{equation}
where $H^t$ is the familiar Hamiltonian function and $H^x$ is its
flux.

The Witten-Zuckerman current $2$-form $\omega^{\mu}$ is derived
from the variational principle underlying the equations of motion.
In the case of the Lagrangian (\ref{lag1}) we need to consider its
first variation assuming the equation of motion (\ref{potkdv})
and its Jacobi equation
\begin{equation}
\delta \phi_{t} + 6  \, \phi_{x} \, \delta \phi_x +  \delta
\phi_{xxx}   = 0. \label{jacobi}
\end{equation}
Then the first variation of the Lagrangian reduces to a
conservation law
\begin{equation}
\delta {\cal L}  = \alpha^{t}_{\; ,t} + \alpha^{x}_{\; ,x}
\end{equation}
where $\alpha^\mu$ is a $1$-form. For the Lagrangian (\ref{lag1})
we find
\begin{equation}    \begin{array}{lll}
\alpha^{t} & = &   \frac{\textstyle {\partial {\cal L}
    } }   {\textstyle {\partial \phi_t} } \,  \delta \phi \; = \;
\frac{1}{2} \, \phi_{x} \, \delta \phi , \\[4mm] \alpha^{x} & = &
\frac{\textstyle {\partial {\cal L}
  } }               {\textstyle {\partial \phi_x} }  \delta \phi
 + \frac{\textstyle {\partial  {\cal L}  } }
 {\textstyle {\partial \phi_{xx}} } \delta \phi_x
-  \left( \frac{\textstyle {\partial  {\cal L}  } }
 {\textstyle {\partial \phi_{xx}} } \right)_x  \delta \phi + ...
 \nonumber \\[4mm]
 & =  & \left( \frac{1}{2} \, \phi_t + 3 \, \phi_x^{\;2} +
\phi_{xxx}  \right) \delta \phi - \phi_{xx} \, \delta \phi_{x}.
\end{array}
\label{varfirst}
\end{equation}
and the symplectic current $2$-form $\omega^\mu$ is given by
eq.(\ref{alphao}). We find that
\begin{eqnarray}
\omega^{t} & = &  \frac{1}{2} \, \delta \phi_{x} \wedge \delta
\phi \label{om1t}   \\ \omega^{x} & = &  3  \, \phi_x \, \delta
\phi_x \wedge \delta \phi + \delta \phi_{x} \wedge \delta
\phi_{xx}+ \frac{1}{2} \, \delta \phi_{xxx} \wedge \delta \phi
\label{om1x}
\end{eqnarray}
where we have used the Jacobi equation (\ref{jacobi}). Note that
the freedom of adding an arbitrary divergence term to the
Lagrangian disappears in the Witten-Zuckerman symplectic $2$-form.
We can check that eq.(\ref{om1x}) remains invariant under the
replacement of $-\frac{1}{2} \phi_{xx}^{\;\;2}$ in the Lagrangian
(\ref{lag1}) by its divergence equivalent $\frac{1}{2} \phi_x
\phi_{xxx}$.

It is obvious that both components of $\omega^\mu$ satisfy the
closure property (\ref{wz2fclosed}) but the check of conservation
law (\ref{wz2fcons}) requires use of the Jacobi equation
(\ref{jacobi}) again. Finally, we need to check Hamilton's
equations in the covariant symplectic form of
eqs.(\ref{symhameqcov}). We have already checked the $t$ component
in section \ref{sec-symp}. Using (\ref{contraction}) and its
generalization to higher derivatives, we find that the contraction
of (\ref{om1x}) with (\ref{vectorfield}) yields $\delta H_1^x$
where
\begin{equation}
H_1^x = \frac{9}{2} \, \phi_{x}^{\;4} + 3 \, \phi_x^{\;2}
\phi_{xxx} - 6 \, \phi_x \, \phi_{xx}^{\;\;2} - \phi_{xx}
\phi_{4x} + \frac{1}{2} \, \phi_{xxx}^{\;\;\;2} \label{h1flux}
\end{equation}
is the familiar flux of $H_1$.

We conclude by remarking again that the time component of the
Witten-Zuckerman $2$-form (\ref{om1t}) is precisely the symplectic
$2$-form (\ref{om1}) obtained from Dirac's theory of constraints.

\section{Pavlov's new KdV Lagrangian}

We have derived the first Hamiltonian operator for KdV as the
Dirac bracket for the degenerate Lagrangian (\ref{lag1}) and showed
that it can be obtained from the covariant Witten-Zuckerman theory
as well. But now we know that there is also a second Hamiltonian
operator for KdV. So the question naturally arises as to ``What is
the Lagrangian for which Magri's Hamiltonian operator
(\ref{magri}) is the Dirac bracket?"

I have asked this question, thinking that
it is a rhetorical question, in several talks that I gave on integrable
systems. Last time I did so, at NEEDS 2000 in G\"okova, my friend Pavlov
\cite{maxim} got up to say that there are indeed many Lagrangians
for KdV and showed me that
\begin{equation}
{\cal L}_{-1} = \left( \phi_{x}^{\;2} +  \frac{1}{2}\, \phi_{xxx}
\right) \phi_t +  \left( \frac{5}{2} \, \phi_{x}^{\;4}  - 5 \,
\phi_x \, \phi_{xx}^{\;\;2} + \frac{1}{2} \, \phi_x^{\,3} \right)
\label{lag2}
\end{equation}
is another Lagrangian for KdV. Indeed one can verify it directly.
But the point is that this Lagrangian is not just an inspired
guess. There is a direct derivation of (\ref{lag2}) which
opens up the flood gates to an infinite family of Lagrangians for
KdV as the subscript $-1$ indicates.\footnote{I have since
received email from Z. Popowicz who has written a computer
program to generate Lagrangian after Lagrangian for KdV. Because
he is a computer expert he couldn't resist the temptation to
invent his own notation. Unfortunately, I found his notation
absolutely impossible to decipher. Fortunately, there is no need to
do so. In the next paragraph we shall present the derivation of all
these Lagrangians.}

The derivation of Pavlov's new Lagrangian for KdV is based the
Lenard recursion relation (\ref{lenard}). To understand this, let
us first go back to the classical Lagrangian (\ref{lag1}) and see
how, in retrospect, we would derive it. For $n=1$ the recursion
relation (\ref{lenard}) can be written in the form
\begin{equation}
u_t = (\phi_{x} )_t = J_{0} \, \delta_u H_1 = \delta_{\phi} \,
H_1(u=\phi_x) \label{method}
\end{equation}
in view of eqs.(\ref{gardner}) and (\ref{clebsch}). Indeed, in the
Lagrangian (\ref{lag1}) terms that do not depend on the velocity
precisely make up $H_1$. In order to obtain the full Lagrangian
(\ref{lag1}) all that remains to be done is to rewrite the left
hand side of (\ref{method}), namely $\phi_{tx}$ in variational
form. In complete analogy, for Pavlov's Lagrangian terms which
do not involve the velocity consist of $H_2$ given by (\ref{h2}),
the next conserved Hamiltonian function in the KdV hierarchy.
Then we go back to the Lenard recursion relation and write it for $n=2$
\begin{equation} \begin{array}{llllll}
 &J_0 \, \delta_u H_2 & = & J_1 \, \delta_u H_1 & = & {\cal R} \,
J_0 \, \delta_u H_1 \\ = & \delta_\phi H_2  & &   & = &  {\cal R}
\, \delta_\phi H_1 = {\cal R} \, \phi_{tx} \end{array}
\label{derive}
\end{equation}
and therefore for Pavlov's Lagrangian the factor in front of
$\phi_t$ is simply obtained by writing the action of the recursion
operator on $\phi_{tx}$ in variational form. From this
construction it is manifest that starting with $H_n$ all we need
to do in order to obtain a new Lagrangian ${\cal L}_{-n}$ for KdV
is to write ${\cal R}^n \, \phi_{tx} $ in variational form. Hence
we have

{\bf Theorem:} {\it For every Hamiltonian function in the KdV
hierarchy, there exists a degenerate Lagrangian that yields KdV as
its Euler-Lagrange equation}.

The number of Lagrangians for KdV is therefore infinite in number.
However, here we must note that while KdV will be an extremum for
the first variation of all these Lagrangians, the Euler equations
require something weaker. Namely, the $n^{th}$ power of the action
of the recursion operator on KdV should vanish.

Finally, we note that
the Lenard recursion operator (\ref{recursionop}) is non-local and
its repeated application on $\phi_{tx}$ will require the introduction
of non-local terms in higher Lagrangians. This is not a problem
because the original KdV Lagrangian is itself non-local! The
Lagrangian (\ref{lag1}) is not expressible in terms
of the original variable $u$ but by its potential (\ref{clebsch}), or
$\phi = ( d / d x)^{-1} \, u$. Higher Lagrangians for KdV can
be written in local form by introducing potentials for the
Clebsch potential according to $\phi = \psi_x$ etc.

\section{New symplectic structure of KdV}

Every new Lagrangian for KdV will give rise to a new symplectic
structure. This can be obtained either through Dirac's theory of
constraints, or directly through the Witten-Zuckerman theory. We
shall now briefly discuss both of these procedures for Pavlov's
Lagrangian (\ref{lag2}). The result will be a new symplectic
structure for KdV. It is not the one we were questing after: The
new symplectic $2$-form is not the inverse of Magri's Hamiltonian
operator.

\subsection{Dirac constraint analysis}

Pavlov's Lagrangian (\ref{lag2}) is degenerate because it is linear in
the velocity $\phi_t$. Its Hessian (\ref{hessian}) vanishes identically.
So we need to apply Dirac's theory of constraints in order to cast it into
Hamiltonian form. The constraint obtained from the definition of momentum
that follows from the Lagrangian (\ref{lag2}) is given by
\begin{equation}
\Phi = \pi -  \phi_{x}^{\;2} - \frac{1}{2}\, \phi_{xxx}
\label{momenta2}
\end{equation}
and it is second class because
\begin {equation}
 \{\Phi(x), \Phi(y)\}_P   = \frac{1}{2} \delta_{xxx} (y-x)
 - \frac{1}{2}  \delta_{yyy} (x-y)
 + 2 \, \phi_x \,  \delta_{x} (y-x) -
2 \, \phi_y \, \delta_{y} (x-y) \label{pbconst3}
\end{equation}
does not vanish modulo (\ref{momenta2}). The symplectic $2$-form,
obtained through (\ref{defomega}) and (\ref{pbceqinv}) is given by
\begin {equation}
\omega_{-1}  =  \frac{1}{2} \, \delta \phi_{xxx} \wedge \delta
\phi + 2 \, \phi_x \, \delta \phi_{x} \wedge \delta \phi
\label{om2}
\end{equation}
which is manifestly a closed $2$-form. Now Hamilton's equations
(\ref{symhameq}) are satisfied with $H_2$ on the right hand side.

In order to obtain the Hamiltonian operator that corresponds to
this new symplectic structure for KdV we should invert the Poisson
bracket of the constraints (\ref{pbconst3}) through the definition
(\ref{mik}). Here we encounter another surprise because the
equation satisfied by the two-point function $J_{-1}(x,y)$ that
will lead to the new Hamiltonian operator through the definition
in eq.(\ref{defhop}) is given by
\begin {equation}
\frac{d^3}{d x^3} J_{-1}(x,y) + 4 \, \phi_x \frac{d}{d x}
J_{-1}(x,y) + 2 \, \phi_{xx} \, J_{-1}(x,y) = \delta (x-y)
\label{surprise}
\end{equation}
which is precisely the equation to be solved for the inverse of
Magri's second Hamiltonian operator (\ref{magri}). The inverse of
Magri's operator is not known. One could write it as an infinite
series which, however, does not appear to be tractable. So we are
unable to write the minus first Hamiltonian operator explicitly
but we have arrived at a curious result. Namely, {\it the inverse
Magri's Hamiltonian operator (\ref{magri}) is also a Hamiltonian
operator for KdV.}

\subsection{Witten-Zuckerman analysis}

The covariant symplectic $2$-form for Pavlov's Lagrangian is
derived along the same lines as in section \ref{sec-wz}. Here we
shall only record the final result. The time component of the
Witten-Zuckerman symplectic $2$-form is the same as the expression
in eq.(\ref{om2}) obtained from Dirac's theory. Now, however, we
also have its flux
\begin{eqnarray}
\omega_{-1}^x & = & (12 \,\phi_x^{\;2} + 5 \, \phi_{xxx}  ) \,
\delta \phi_{x} \wedge \delta \phi - 6  \, \phi_{xx} \, \delta
\phi_{xx} \wedge \delta \phi + 5 \, \phi_{x} \, \delta \phi_{xxx}
\wedge \delta \phi \label{om2flux} \\ & &  - \frac{1}{2}  \,
\delta \phi_{5x} \wedge \delta \phi - 4 \, \phi_x \, \delta
\phi_{xx} \wedge \delta \phi_x - \frac{3}{2} \,  \delta \phi_{4x}
\wedge \delta \phi_x + \frac{1}{2} \delta \phi_{xxx} \wedge \delta
\phi_{xx}. \nonumber
\end{eqnarray}
The check that Hamiltonian equations in the covariant form of
eqs.(\ref{symhameqcov}) are satisfied with this symplectic
$2$-form and the second Hamiltonian function $H_2$ (\ref{h2}) is
lengthy but straight-forward. For the space component of the
Witten-Zuckerman $2$-form it is even lengthier to check that these
equations are satisfied with (\ref{om2flux}) and
\begin{eqnarray}
H_{2}^x & = & 12 \,\phi_x^{\;5} + 10 \, \phi_{x}^{\;3} \phi_{xxx}
- 45  \, \phi_{x}^{\;2}  \, \phi_{xx}^{\;\;2} + 8 \, \phi_{x} \,
\phi_{xxx}^{\;\;\;2} + 5 \, \phi_{xx}^{\;\;2} \, \phi_{xxx}
\nonumber \\ & & + \phi_{xxx}  \phi_{5x} - \frac{1}{2} \,
\phi_{4x}^{\;\;2},  \label{h2flux}
\end{eqnarray}
the flux of $H_2$ on the right hand side of (\ref{symhameqcov}).

\section{Conclusion}

The Lagrangian approach to integrable non-linear evolution
equations is one where we can derive everything from first
principles. We had earlier used it for various equations ranging
from dispersive water waves \cite{nn} to Monge-Amp\`ere equations
\cite{rma}, \cite{cma} in differential geometry. Here we have
tried to illustrate this approach using the first Hamiltonian
structure of KdV. Starting with the classical Lagrangian for KdV
we applied Dirac's theory of constraints to arrive at its first
Hamiltonian operator. We showed that the covariant
Witten-Zuckerman theory of symplectic structure not only yields
the symplectic $2$-form obtained from Dirac's theory but also
gives us the flux component of the symplectic $2$-form. Then we
applied these approaches to Pavlov's new Lagrangian and arrived at
a new symplectic structure for KdV. We have arrived at the
remarkable result that the inverse Magri's Hamiltonian operator
is also a Hamiltonian operator for KdV. Pavlov's
procedure for constructing new Lagrangians is applicable to all
conserved Hamiltonians of which KdV has infinitely many. Therefore
there will be infinitely many symplectic structures for KdV.

\section{Acknowledgement}

Special thanks and congratulations are due to Professors Henrik
Aratyn and Alexander Sorin for organizing such a stimulating NATO
ARW.


\begin{thebibliography}{99}

\bibitem{magri} Magri F 1978 J. Math. Phys. {\bf 19} 1156 \\
Magri F 1980 in {\it Nonlinear Evolution Equations and Dynamical
Systems} Boiti M, Pempinelli F and Soliani G editors Lecture Notes
in Phys., {\bf 120}, Springer, New York, p. 233 \\ Magri F, Morosi
C and Tondo G 1988 Comm. Math. Phys. {\bf 115} 457
\bibitem{dirac} Dirac P A M 1964 {\it Lectures on Quantum Mechanics}
Belfer Graduate School of Science Monographs series 2, New York \\
Hanson A, Regge T and Teitelboim C 1976 Acad. Naz. Lincei (Rome)
\\ Sundermeyer K 1982 {\it Constrained Dynamics}, Lecture Notes in
Physics Vol. 169 Springer Verlag \\ Olver P J 1986 {\it Applications
of Lie Groups to Differential Equations}, Graduate Texts in
Mathematics Vol. 107 Springer Verlag \\ Kosmann-Schwarzbach Y 1986
{\it G\'eom\'etrie de syst\'emes Bihamilton\-iens}, Univ. of
Montreal Press No.{\bf 102} 185
\bibitem{irene} Dorfman I Ya 1993 {\it Dirac Structures} J. Wiley \& Sons
\bibitem{yavuz} Nutku Y 1983 J. Phys. A: Math. and Gen. {\bf 16} 4195
\bibitem{witten} Witten E 1986 Nuclear Physics {\bf B 276} 291 \\
Crnkovi\'{c} C and Witten E 1986 in {\it Three Hundred Years
of Gravitation}, Hawking S W and Israel W editors
\bibitem{zuck} Zuckerman G J 1986 in {\it Mathematical Aspects of String Theory} Yau S T
editor, World Scientific
\bibitem{diracadm} Dirac P A M 1958 Proc. Roy. Soc. (London) {\bf A 246} 217
\bibitem{mcf}  Macfarlane A J 1982 CERN preprint TH 3289
\bibitem{ynkdv} Nutku Y 1984 J. Math. Phys. {\bf 26} 2007
\bibitem{gardner} Gardner C S 1971 J. Math. Phys. {\bf 12} 1548
\bibitem{zf}Zakharov V E and Fadeev L D 1971 Funct. Anal. Appl. {\bf 5} 18
\bibitem{capg} Galvao C A P 1994 Private communication.
\bibitem{fokas} Santini P M and Fokas A S 1988 Commun. Math. Phys. {\bf 115}
375  \\ Fokas A S and Santini P M 1988 Commun. Math. Phys. {\bf
116} 449 \\ Dorfman I Ya and Fokas A S 1992 J. Math. Phys. {\bf
33} 2504
\bibitem{maxim} Pavlov M 2000 Private communication.
\bibitem{nn} Neyzi F and Y Nutku 1987 J. Math. Phys. {\bf 28} 1499
\bibitem{rma} Nutku Y 1996 J. Phys. A {\bf 29}  3257
\bibitem{cma} Nutku Y 2000 Phys. Lett. A {\bf 268} 293

\end{thebibliography}
\end{document}